\documentstyle[12pt]{article}
\def\graphic #1#2#3#4#5{

    \noindent
    \centerline{\hrulefill}
    \leftline{\hbox to#1{\special{anisoscale #3, #1 #2}\hfil}}
    \vspace*{#2} \relax
    \vskip -3.9 cm
    \hskip 4.8 cm
    {\large \bf Universidade do Estado do Rio de Janeiro }
    \newline

    \vskip -0.25 cm
    \hskip 7.5 cm
    {\large \bf Instituto de F{\'\i}sica }

    \vskip 1 cm
    \hskip 7.5 cm
    {\large Phys-Pub #4 }

    \hskip 7.5 cm
    {\large Preprint}

    \hskip 7.5 cm
    {\large #5 }

    \medskip
    \noindent
    \hrulefill

    \vskip 2.9 cm
    }
\def\({\c c}
\def\|{\'\i}

\def\dgraphic #1#2#3{
    \centerline{\hbox to#1{\special{anisoscale #3, #1 #2}\hfil}} 
    \vspace*{#2} \relax         
    }
\def\dtwographic #1#2#3#4{
    \centerline{\hbox to#1{\special{anisoscale #3, #1 #2}\hfil}
                \hbox to#1{\special{anisoscale #4, #1 #2}\hfil}}
    \vspace*{#2} \relax
    }
\def\dthreegraphic #1#2#3#4#5{
    \centerline{\hbox to#1{\special{anisoscale #3, #1 #2}\hfil}
                \hbox to#1{\special{anisoscale #4, #1 #2}\hfil}
                \hbox to#1{\special{anisoscale #5, #1 #2}\hfil}}
    \vspace*{#2} \relax
}
\def\x{\mbox{\tiny x}}
\begin{document}
\hspace\parindent
\thispagestyle{empty}

\graphic{2 in}{1.6 in}{uerj}{01/96}{January 1996}
\centerline{\LARGE \bf Poincar\'e Sections of Hamiltonian Systems}


\bigskip
\bigskip
\centerline{\large
E.S.Cheb-Terrab\footnote{Departamento de F\|sica Te\'orica, IF-UERJ.
URL: http://dft.if.uerj.br \newline
\hspace*{.65cm}E-mail: terrab@vmesa.uerj.br, \newline
\hspace*{.65cm}E-mail: henrique@vmesa.uerj.br}
and
H.P. de Oliveira\footnotemark[1]
}

\bigskip
\bigskip
\bigskip
\begin{abstract}
A set of Maplev R.3 software routines, for plotting 2D/3D projections of
Poincar\'e surfaces-of-section of Hamiltonian dynamical systems, is
presented. The package consists of a plotting-command plus a set of
facility-commands for a quick setup of the Hamilton equations of motion,
initial conditions for numerical experiments, and for the zooming of plots.
\end{abstract}

\bigskip
\centerline{ \underline{\hspace{6.5 cm}} }

\medskip
%
%
\centerline{ {\bf (Submitted to Computer Physics Communications)} }

\newpage
\bigskip
\hspace{1pc}
{\bf PROGRAM SUMMARY}
\bigskip

\begin{footnotesize}
\noindent
{\em Title of the software package:} Poincar\'e.   \\[10pt]
{\em Catalogue number:} (supplied by Elsevier)                \\[10pt]
{\em Software obtainable from:} CPC Program Library, Queen's
University of Belfast, N. Ireland (see application form in this issue)
\\[10pt]
{\em Licensing provisions:} none  \\[10pt]
{\em Operating systems under which the program has been tested:}
UNIX systems, Macintosh, DOS (AT 386, 486 and Pentium based) systems,
DEC VMS, IBM CMS.                        \\[10pt]
{\em Programming language used:} {\bf Maple V} Release 3. \\[10pt]
{\em Memory required to execute with typical data:}  8 Megabytes. \\[10pt]
{\em No. of lines in distributed program, including On-Line Help,
etc.:} 1381.                                                   \\[10pt]
{\em Nature of mathematical problem}\\
Computation and plotting of 2D/3D projections of Poincar\'e surfaces-of-section
of Hamiltonian systems.
   \\[10pt]
{\em Methods of solution}\\
A $4^{\mbox{\scriptsize th}}$ order Runge-Kutta method with optional stepsize
and number of iterations is used. However, it is possible
to indicate any user-method to be used in the integration scheme.
   \\[10pt]
{\em Restrictions concerning the complexity of the problem}\\ Besides the
inherent restrictions of the Runge-Kutta method, this first version of
the package does not makes use of adaptative stepsize control.
   \\[10pt]
{\em Typical running time}\\
It depends strongly on the surface-of-section to be plotted. With a Pentium-90
PC (32 Mb. RAM), fast plots usually take from a few seconds to a few minutes.
On the other extreme, in an example considered in this paper, a
surface-of-section with 10,000 points and an energy threshold $\approx 10
^{-8}$
took 35 minutes.
   \\[10pt]
{\em Unusual features of the program}\\
This package provides easy-to-use software tools for plottings 2D/3D
projections of Poincar\'e sur\-fa\-ces-of-section of Hamiltonians systems. The
speed at which the returned plots are calculated is adjustable, in connection
with their accuracy. This feature permits alternatively searching for, say,
``first order" phenomena at remarkable high speed, or, say, ``high order"
detailed 2D/3D projections displaying ``islands" and the inner structure of a
surface-of-section, as desired. The 2D intersection plane over which the
surface-of-section is plotted can be any one of the coordinate planes of the
phase space, and can be shifted in the positive and negative directions. The
package also provides routines for setting large sets of initial conditions
for numerical experiments in seconds. The implementation in a symbolic
computing environment allows for combined symbolic/numerical studies.
\end{footnotesize}
\newpage
\hspace{1pc}
{\bf LONG WRITE-UP}

\section*{Introduction}


The Poincar\'e surface-of-section method has become a popular technique for
analyzing weakly perturbed Hamiltonian systems\cite{lichtenberg}. People
working in correlated areas usually prepare numerical routines, for instance
in Fortran or C, and use the computational environment to find the solution
curves of the model, and to build the corresponding surfaces-of-section
plots. In parallel, symbolic systems now present a satisfactory performance
when computing with hardware floating-point numbers, while offering a
comfortable computational environment for realizing symbolic mathematical
studies and manipulating plots.

Related to these points, this paper presents a set of software routines,
implemented in MapleV R.3, for plotting Poincar\'e surfaces-of-section,
exploiting the MapleV routines for working with hardware floats and with
plots. In preparing such a package, our intention was:
\begin{itemize}
\item to provide easy-to-use software tools for fast plottings of Poincar\'e
sections of Hamiltonian dynamical systems;
\item to make these tools flexible enough to feature sections over any
possible 2D/3D submanifold of the corresponding phase space (including the
time as a possible variable);
\item to allow for combined symbolic/numerical studies by implementing such
software tools in a symbolic computing environment.
\end{itemize}

The exposition is organized as follows. In Sec.\ref{theory}, the basic ideas
concerning the surface-of-section method are briefly reviewed. In
Sec.\ref{package}, a summary of the package's commands is presented, followed
by a detailed description\footnote{Aside from this, the package itself
contains an {\it On-Line} help in standard Maple format which can be viewed as
the User's manual for all the routines.} of its most relevant commands, mainly
{\bf poincare}, for building the 2D/3D plots of Poincar\'e surface-of-section
projections, and {\bf gin}, for generating sets of initial conditions for
numerical experiments. Sec.\ref{examples} contains a brief illustration of how
the new commands work in three selected examples that can also be regarded as
simple {\it input/output} tests. They are: the Toda lattice\cite{lichtenberg},
the H\'enon-Heiles Hamiltonian\cite{henonheiles}, and a numerical study, in
the context of general relativity, which appeared in a recent
publication\cite{calzetta2}. Finally, the Conclusions contain a brief
discussion about this work and its possible extensions.

\section{The {\it Poincar\'e} surface-of-section method}
\label{theory}
The Poincar\'e surface-of-section method has become popular in the last
decades in connection with the KAM theorem for weakly perturbed (originally
integrable) Hamiltonian systems.


An insight of the ideas underlying this method can be obtained by considering,
for instance, a conservative system with two degrees of freedom. The physical
trajectories lie on the three-dimensional energy surface $H(p_1,p_2,q_1,q_2) =
H_0$ and, if the motion is bounded, after a suitable time interval, the
solution curves will repeatedly intersect any given embedded 2D-plane; for
instance, the $q_2=constant$, $(p_1,q_1)$ plane. If another constant of motion
involving $p_1$ and $q_1$ exists (then the system is integrable), it is
possible to use it to express $p_1=p_1(q_1)$, i.e., the intersection points
will lie on a smooth curve. The enclosed area will be an integral invariant
and, as time flows, these smooth curves will draw surfaces in the phase space.

For weakly perturbed systems, these surfaces of solution curves of regular
motion (KAM surfaces) continue to exist for most initial conditions, breaking
their topology near resonances to form ``islands" chains. Within these
islands, the topology is broken again to other chains and so on. Generally
speaking, the KAM surfaces isolate thin layers of stochasticity and, as the
perturbation strength is increased, transitions between layers merge, and the
KAM surfaces progressively disappear resulting in complete stochastic motion.

In this manner, the plotting of the intersection points of the solution curves
with a given 2D-plane (the {\it surface-of-section}) permits the study of the
existence of non-obvious constants of motion (isolating integrals in the
context of Hamilton-Jacobi theory), local stability, transition from ordered
to stochastic motion, and many other interesting properties. We recall that no
general procedure for determining the integrability of an arbitrary
Hamiltonian system, or even the number of such isolating integrals, has yet
been found. As a consequence, the plotting of these surfaces-of-section plays
an important role not only in numerical studies, but also in checking the
consistency of analytical results obtained using perturbative methods.

In the multidimensional case, the Poincar\'e surface-of-section (PS) has
dimensionality $2N$-$2$. Although the case $N \ge 3$ presents some subtleties
(Arnold diffusion etc.), for trajectories which are exactly separable in the
$(p_i,q_i)$ coordinates, the intersection points of the solution curves with
the corresponding $(p_i,q_i)$ plane also lie on a smooth curve. We will
denominate {\it two-dimensional surface-of section} (2PS) the plotting of
these intersection points over a given 2D-plane of the phase space. In the
general case the intersection points may not lie in such smooth curves, but
they still fill an annulus of finite area, and the thickness can be related to
the nearness concerning exact separability in the related $(p_i,q_i)$
coordinates.

It is possible to extend the concept of 2D surface-of-section to {\it 3D
space-of-section} (3PS), which will contain 3D projections of the solution
curves. The plotting of the 2PS embedded in a related 3PS may give
interesting detailed information about the behavior of a given model.

\section{The {\it Poincar\'e} package}
\label{package}

Basically, the {\it Poincar\'e} package consists of a plotting-command, {\bf
poincare}, for the 2D/3D plotting of the corresponding projections of
surfaces-of-section, plus a set of facility-com\-mands for a quick setup of
the Hamilton equations of motion, initial conditions for numerical
experiments, and for the zooming of plots. All plots are built by numerically
integrating Hamilton's equations for a given set of initial conditions. Once
the plot is realized, all the Maple facilities for handling plots are
available. Complementary symbolic studies in the same environment can be
developed with the commands of the {\it PDEtools} package\cite{PDEtools} for
the analytical solving of systems of ODE's, scalar PDE's (the corresponding
Hamilton-Jacobi equation), and for changing variables.

\subsection*{\it Summary}

A brief review of the commands of the package is as follows\footnote{This
subsection and the next one may contain some information already presented in
the previous sections; this was necessary to produce a complete-cut description
of
the package.}:

\begin{itemize}

\item {\bf hameqs} receives a Hamiltonian and returns Hamilton's equations
and a list with the p's and q's involved;

\item {\bf poincare} receives a Hamiltonian, a time range and initial
conditions, and returns a 2D plot of a 2PS; or a 3D plot of 2PS embedded in
a related 3PS. In both the 2D and 3D plots, the 2D plane over which the 2PS is
plotted can be any of the coordinate planes of the phase space, and can
optionally be shifted in the positive or negative directions;

\item {\bf gin} generates a set of lists of initial conditions for plotting a
2PS/3PS, from a given incomplete set of fixed values and/or ranges for the
$p$'s, $q$'s and the energy. This command speeds-up the time-expensive task of
setting appropriate initial conditions for performing numerical experiments;

\item {\bf zoom} allows for changing the ranges of the display of a given
2D/3D plot without having to recalculate it again, thus saving time and memory
resources.

\end{itemize}

\subsection*{\it Description}

A complete description of the {\it Poincar\'e} package's commands is found in
the On-Line help. Therefore, a detailed description will be given here only
for the most relevant routines, namely {\bf poincare} and {\bf gin}. Some
commented {\it input/output} examples can be found in Sec.\ref{examples}.

\subsection{Command name: {\bf poincare}}
\label{poincare}

\noindent {\it Feature:} plot 2D/3D projections of the Poincar\'e
surface-of-section of a given Hamiltonian system.

\noindent {\it Calling sequence\footnote{In what follows, the {\it
input} can be recognized by the Maple prompt \verb->-.}:}
\begin{verbatim}
> poincare(H, t=a..b, ics, optional_parameters);
\end{verbatim}

\noindent
{\it Parameters:}

\noindent
\begin{tabular}{ll}
\verb-H-           & - any algebraic expression representing the Hamiltonian.
\\
\verb-t=a..b-      & - {\tt t} represents the time and {\tt a..b} is a
                         numerical range.
\\
\verb-ics- & - a set of initial conditions for the $p$'s and $q$'s in the
form: \\
                   & \ \{{\tt {[n1,n2,n3,...], [...],...}\}}, where
                   {\tt [n1,n2,n3,...]} is a list of numbers \\
                   & \ \ representing the
                   initial values for {\tt [t, p1, p2,...pk, q1,q2,...,qk]}.
\end{tabular}

\noindent
{\it Optional Parameters:}

\noindent
\begin{tabular}{ll}
\verb-stepsize=n- & - indication of a positive number representing the size of
                      the step\\
                  & \ \  used in the integration scheme.
\\
\verb-iterations=N- & - indication of a positive integer representing the
                        number of \\
                  & \ \ iterations used in the integration scheme.
\\
\verb-scene=[xi,xj]- & - indication of the variables constituting the 2D-plane
                         of the phase \\
                  & \ \ space where the 2PS is plotted; it is
                     possible to specify\\
                  & \ \ ranges for the plot as in: {\tt
scene=[xi=a..b,xj=c..d]}
\\
\verb-scene=[xi,xj,xk]- & - indication of the 2D-plane and a third variable,
{\tt xk},
                            to be used as \\
                  & \ \ {\it cross-variable} or third axis in plots of
2PS/3PS's,
                         respectively
\\
\verb-shift=s-    & - indication of a number representing a positive or
negative
                      shift\\
                  & \ \ of the intersection plane in the plots of 2Ps/3Ps
\\
\verb-method=procedure- & - indication of a user procedure to be used as
                            integration method.
\\
\verb-3-          & - to obtain a related 3D plot (3PS).
\end{tabular}

\medskip
\noindent
{\it Synopsis:}

The {\bf poincare} command was designed to build either fast but not so
accurate, or slower, as accurate as desired, 2D/3D projection plots of
Poincare sections (see optional arguments below). Instead of analytically
enforcing the Hamiltonian constraint, the value of the energy of each plotted
point is checked against the corresponding $H_0$. As a complement, another
routine, {\bf gin} (see subsection 3.2), was programmed in order to speed-up
the preparation of initial conditions for the numerical experiments.

The returned plots can be manipulated using all the standard Maple facilities
(icon tool-bar) for 2D/3D plots such as reorientation, perspective, etc., and
using the {\bf zoom} command, also part of this package, for zooming
in/out the interesting regions. These facilities usually permit a
detailed visual distinction between the KAM surfaces and the layers of
stochasticity for, say, typical weakly perturbed systems.

In addition to the returned plot, some relevant information related to each
solution curve is displayed on the screen during the calculations. This
information is:
\begin{itemize}
\item the initial value of the Hamiltonian, p's and q's for the curve;
\item the number of intersection points found in the given time interval
(2D plots);
\item the maximum percentile ``energy-deviation" of the intersection points.
\end{itemize}
It is useful to know the number of intersection points since it may be an
indication of how appropriate is the indicated time interval. Concerning the
percentile energy-deviation\footnote{When $H_0=0$, just the maximum absolute
deviation is displayed.}, it is calculated as $\frac{H_0 -
H(point)}{H_0}\,100$. We note that all numerical algorithms will lead to
values of $H$ different from the initial $H_0$, specially in the case of
optional fast plottings. Percentile deviations below $10^{-8}$ are displayed
as $0$. Also, the deviation is calculated for each intersection point, but
only the greatest value is displayed. This information will give an idea of
the accuracy of the plot, and the user will have the option of either
reentering the instruction looking for a more accurate/slower plot or using
his/her own numerical integration algorithm (see
below). In typical situations, the smooth curves can be recognized even with
energy-deviations of the order of $H_0/10$.

\medskip
\noindent
{\it The arguments}

The first argument of a {\bf poincare} calling is the Hamiltonian. Some handy
conventions were adopted to represent the $q$'s and $p$'s. These are: all
$p$'s and $q$'s must appear as $pn$ or $qn$ where n is a positive integer, as
in $p1$, $p2$, etc., and the time dependence need not be explicit, as in $pn$
or $qn$ instead of $qn(t)$ or $pn(t)$. The Hamiltonian is assumed to be time
independent and the number of degrees of freedom\footnote{To avoid useless
large code this number was restricted to 10 (20-dimensional phase space), but
this restriction can easily be removed.} is expected to be $\leq 10$.

The solution curves are calculated within a given time interval specified in
the second argument as {\tt t=a..b}. When this time range leads to no
intersection points, one can use the \verb-3- option, see how far the
intersection plane is from the trajectories, and reenter the instruction
shifting the intersection plane using the \verb-shift- option (see below).

The third argument, a set, may have any number of lists of initial conditions
for the time, {\tt p}'s and {\tt q}'s, corresponding or not to the same
initial value of $H$; they can be generated by the user or by the {\bf gin}
command, as explained in subsection \ref{gin}. The initial conditions must be
inside a set structure as in {\tt \{[n1,n2,n3,...], [...],...\}}, where {\tt
[n1,n2,n3,...]} is a list of numbers representing the initial values for {\tt
[t, p1,...pk, q1,...qk]}.

\medskip
\noindent
{\it The optional arguments}

The optional arguments can be given alone or in conjunction and in any order.

By default, the step interval is {\tt (b-a)/20}, where {\tt a..b} is the range
for {\tt t}. This can be changed by giving the extra argument {\tt stepsize =
n}, where n is a positive number. As the stepsize is decreased, the accuracy
and the smoothness of the integral curves (as well as the time
consumed in the calculations) will increase.

The default numerical algorithm used in the integration scheme is basically
the $4^{\mbox{\scriptsize th}}$ order Runge-Kutta of the {\it DEtools}
package, but this can be changed by giving the extra argument {\tt method =
usermethod}, where \verb-usermethod- should be a numerical integration
algorithm (see the help pages of the {\it DEtools} package). Also, when
requesting a plot, the numerical algorithm can be iterated, as many times as
desired, by giving the extra argument {\tt iterations = N} (default:
iterations = 1).

The default scene for the plots is: the $(p_1,q_1)$ plane, at $q_2=0$ for the
2PS, or the $(p_1,q_1,q_2)$ 3D-submanifold, for the plot of a 2PS embedded in
a 3PS, when the \verb-3- option is indicated. The intersection points
constituting the 2PS are obtained by looking for the sign change of a ``third"
coordinate, here denominated the {\it cross-variable}, by default $q_2$. The
default ranges for the display of a plot are such as to include all the
calculated intersection points in the case of 2D plots; or all calculated
pieces of projections of solution curves plus the intersection 2D-plane, in
the case of 3D plots.

All these defaults for the scene can also be changed. First of all, the
intersection plane over which the 2PS is plotted can be shifted in the
positive or negative directions by indicating {\tt shift=s} (a real number) as
an additional argument. Concerning the 2D/3D submanifolds or the {\it
cross-variable}, they can be changed by giving the extra argument
\verb-scene=[x1,x2]- or \verb-scene=[x1,x2,x3]-, with or without extra ranges
(for displaying a plot of a particular region), as in
\verb-scene=[x1=a..b,...]-. The 2PS will then be plotted over the plane formed
by the first two variables appearing in the right-hand-side, and $x_3$, when
given, will be the {\it cross-variable}; or the third axis when the \verb-3-
option is given as argument too\footnote{It is possible to indicate the time
"t" as the third variable, in which case a convenient mouse-manipulation of
the 3D-plot can display the projection of the curves over each $(q_i,q_j)$
plane. This may be useful to study the bounded/unbounded properties of a given
potential. Furthermore, in the case of a system with 3 degrees of freedom, the
use of the ``3" option with scene=[q1,q2,q3] will render the 3D-plot of the
physical trajectory.}. When ranges are indicated, it is still possible to zoom
in/out the resulting plot, up to the default ranges mentioned above, by using
the {\bf zoom} command.

\noindent
\subsection{Command name: {\bf gin}}
\label{gin}

\noindent {\it Feature:} generates a set of lists of initial conditions
satisfying the Hamiltonian constraint.

\noindent {\it Calling sequence:}
\begin{verbatim}
> gin(H,ic,N);
\end{verbatim}

\noindent
{\it Parameters:}

\noindent
\begin{tabular}{ll}
\verb-H-           & - any algebraic expression representing the Hamiltonian.
\\
\verb-ic- & - a set of single initial conditions for the time $t$, $p$'s, $q$'s
and the energy,
              in the form:\\
                   & \ \ {\tt \{t=X1,p1=X2,..,qn=Xk,energy=X2n+2)\}}, where the
                   {\tt Xi} are numbers or ranges of \\
                   & \ \ numbers, representing initial values for the time,
                   $p$'s, $q$'s, and the energy.
\\
\verb-N-           & - a positive integer representing the number of desired
                   lists of initial conditions
\end{tabular}

\medskip
\noindent
{\it Synopsis:}

The numerical study of a given Hamiltonian system usually requires a lot of
suitable lists of initial conditions (IC's) satisfying the Hamiltonian
constraint (see examples in Sec.\ref{examples}). The {\bf gin} command was thus
proposed as a tool for the fast generation of a set of such lists,
with the appropriate syntax as required by the {\bf poincare} command
explained above.

The first argument received by {\bf gin} is the Hamiltonian. As second
argument, a {\it single} specification of IC's for all but one of
$(t,p's,q's,H_0)$ must be given. If IC's are specified for all these
variables, the command just checks the values against $H$ for consistency. The
IC for a given variable may be a fixed number or a numerical range.

The third argument is the number of requested lists of IC's. Face the request
of say $N$ IC's giving ranges or fixed numbers for all variables but one, say
$x$, the routine proceeds as follows. The received IC's are separated into
numerical ranges and fixed numbers. Each range is then uniformly
divided into $N$ numbers, and $N$ lists are built by taking one number, from
each divided range, together with the received fixed numbers for the other
variables. These lists are sequentially introduced in $H$, resulting in $N$
algebraic equations for $x$. Each equation is then numerically solved, and a
set of $N$ lists of complete IC's satisfying the Hamiltonian constraint, with
values inside the given ranges, is returned. Remarkably, though {\bf gin} is a
very simple command, it plays a fundamental role in speeding-up the numerical
studies.

\section{Examples}
\label{examples}
This section contains a brief illustration of how the routines here presented
work in some selected examples: the Toda lattice, the H\'enon-Heiles
Hamiltonian, and a numerical study, in the context of general relativity,
which appeared in a recent publication\cite{calzetta2}. Although the idea is
just to test the obtainment of some of the related PSs using the package's
commands, we also included, for completeness, brief introductions to each
example as well as short comments about the resulting PS plots.
\subsection{The Toda lattice}
Our first example is the three-particle Toda lattice\cite{lichtenberg}. This
model describes the
motion of three particles of equal mass moving on a ring with exponentially
decreasing interactions. Early numerical experiments\cite{ford}, followed by
analytical results\cite{henon}, showed that the related dynamical system is
integrable. After some manipulations, the original Hamiltonian can be written
as:
\begin{equation}
H= \frac {1}{2}\ (
\,{\it p_1}^{2} + {\it p_2}^{2}\,) +
\frac {1}{24}  \left( \! \,{\rm e}^{ \left( \! \,2
\,{\it q_2} + 2\,\sqrt {3}\,{\it q_1}\, \!  \right) } + {\rm e}^{
 \left( \! \,2\,{\it q_2} - 2\,\sqrt {3}\,{\it q_1}\, \!  \right) }
 + {\rm e}^{(\, - 4\,{\it q_2}\,)}\, \!  \right)  -
{\displaystyle \frac {1}{8}}
\end{equation}
A fast plot of the surfaces-of-section, projected over the $(p,q)$ planes,
with just one initial condition, can be obtained via:
\begin{verbatim}
> H, t=-150..150, {[0,.1,1.4,.1,0]};   # ics: t=0,p1=.1,p2=1.4,q1=.1,q2=0
> poincare(", stepsize=.05,iterations=5);
> poincare("",stepsize=.05,iterations=5,scene=[p2,q2]);
\end{verbatim}
\dtwographic{3in}{2.6in}{f1a}{f1b}
\noindent
\begin{tabular}{p{3in} p{3in}}
{\footnotesize Fig.1.a. 2PS over the $q_2$=$0$ plane, with 127
intersection points lying on smooth curves. $H_0$=$0.99$, $H$-deviation=$5\,
\mbox{\tiny{x}}10^{-6}\,\%$. Time: $73 s.$}
&
{\footnotesize Fig.1.b. 2PS over the $q_1$=$0$ plane with 146 intersection
points. Time: $76 s.$
}
\end{tabular}

\smallskip
\noindent Generally speaking, it is interesting to have the 2PS projected over
all the $(p_i,q_i)$ planes of the phase space, since the existence of smooth
patterns in all of them is an indication of the possible integrability of the
system.

A Poincar\'e space-of-section corresponding to Fig.1.a can be manipulated
using the mouse to obtain the following illustrative perspectives\footnote{In
what follows, the values of $\theta$ and $\phi$ mentioned in the figures
correspond to the Maple values for the plots.}:
\begin{verbatim}
> poincare(H,t=-100..100, {[0,.1,1.4,.1,0]},stepsize=.1,iterations=4,
>          scene=[p1=-1.5..1.5,q1=-1.5..1.5,q2=-1.2..1.3],3);
\end{verbatim}
\dtwographic{3in}{2.45in}{f2a}{f2b}
\noindent
\begin{tabular}{p{3in} p{3in}}
{\footnotesize Fig.2.a. 3PS projected at $\theta$=$-20$, $\phi$=$75$,
showing a KAM surface of regular trajectories. Time: $41 s.$}
&
{\footnotesize Fig.2.b. A plane projection of the 3PS shows how the
intersection
points are joined outside the 2PS.}
\end{tabular}

\smallskip
Another indication of the integrability of the system is that regular curves
exist whatever the value of $H_0$. As an example of this, a surface-of-section
(one solution curve) and a related 3PS, at $H_0=256$, can be built as follows:
\begin{verbatim}
> ics2 := gin(H,{t=0,p2=22,q1=0,q2=0,energy=256},1):
> poincare(H,t=-50..50,ics2,stepsize=.005,iterations=4,scene=[p2,q2]);
> poincare(H,t=0..20,ics2,stepsize=.01,iterations=4,scene=[p2,q2,q1],3);
\end{verbatim}
\dtwographic{3in}{2.25in}{f3a}{f3b}

\noindent
\begin{tabular}{p{3in} p{3in}}
{\footnotesize Fig.3.a Smooth curves on the 2PS, $q_1$=$0$ plane.
$H_0$=$256$, $H$-deviation=$3\,\mbox{\tiny{x}}10^{-4}\,\%$, 342 points. Time:
$265 s.$}
&
{\footnotesize Fig.3.b.
3PS corresponding to Fig.3.a., $\theta$=$100$, $\phi$=$40$, displaying a KAM
surface constituted by just one regular curve. Time: $40 s.$}
\end{tabular}

\subsection{H\'enon-Heiles Hamiltonian}

The H\'enon-Heiles Hamiltonian produces one of the most famous and studied
surfaces-of-section. The corresponding Hamiltonian can be obtained by
expanding the Toda Hamiltonian to cubic terms in $q_1$ and $q_2$, and is given
by:
\begin{equation}
H=\frac {1}{2}\, \left( p_1^{2} + p_2^{2} + q_1^{2} + q_2^{2} \right) +
q_1^{2}\,q2 - \frac{1}{3}\ q_2^{3},
\end{equation}
This model can be related to the motion of a star in a cylindrically symmetric
gravitational galactic potential, is not integrable, and the phase space is
bounded only if the energy is less than $1/6$. Well-studied
surfaces-of-section, presented in several treatises of chaos, with $H_0$
equal to {$1/24,\, 1/18,\, 1/12,\, 1/8,\, 1/7,$ and $1/6$, are obtained here
using the {\bf gin} and {\bf poincare} commands as follows. To start with, six
sets, related to each value of $H_0$ respectively, with three different initial
conditions each, are generated via:
\begin{verbatim}
> for h in [1/24,1/18,1/12,1/8,1/7,1/6] do
>   ics[h] := gin(H,{t=0,p2=0.1,q2=-0.2..0.2,q1=-0.2..-0.1,energy=h},3)
> od:
\end{verbatim}
After that, surfaces-of-section with around 550 points, calculated in
approximately 5 minutes each, with percentile $H$-deviations $\approx
10^{-7}\,\%$, can be obtained via:

\begin{verbatim}
> for h in [1/24,1/18,1/12,1/8,1/7,1/6] do
>   poincare(H,t=-300..300,ics[h],
>            stepsize=.05,iterations=3,scene=[p2=-.5..0.5,q2=-.5..0.5]);
> od:
\end{verbatim}
\dthreegraphic{2in}{2in}{f4a}{f4b}{f4c}

\noindent
\begin{tabular}{p{2in} p{2in} p{2in}}
{\centerline{\footnotesize Fig.4.a. $H_0=1/24$}}
&
{\centerline{\footnotesize Fig.4.b. $H_0=1/18$}}
&
{\centerline{\footnotesize Fig.4.c. $H_0=1/12$}}
\end{tabular}

\dthreegraphic{2in}{2in}{f4d}{f4e}{f4f}

\noindent
\begin{tabular}{p{2in} p{2in} p{2in}}
{\centerline{\footnotesize Fig.4.d. $H_0=1/8$}}
&
{\centerline{\footnotesize Fig.4.e. $H_0=1/7$}}
&
{\centerline{\footnotesize Fig.4.f. $H_0=1/6$}}
\end{tabular}

\smallskip \noindent The figures above reflect the progressive disintegration
of the KAM surfaces, occurring with the increase of $H_0$ up to 1/6. In the
plots for $H_0$ equal to 1/24 and 1/12, invariant curves apparently exist
everywhere, but this is not strictly correct. In fact, the model is not
integrable, as is reflected by the sequence of figures, and thin resonance
layers of stochasticity are densely distributed throughout the 2PS, even for
small $H_0$.

Another interesting parameter is given by the relevant number of calculations
involved in the building of each of Figs.4.(a,b,c,d,e,f): $600/0.05$=$12,000$
points, iterated $4 \x 3$=$12$ times each; that is, 144,000 calculations in
five minutes, or $480$ calculations per second\footnote{The number of digits
carried in floats is $16$; also, the true {\it calculations per second} is
greater than 480: part of the five minutes were dedicated to determine
the intersection points and the maximum percentile deviation.}.

Concluding this example, an instructive test is to compare the regularity of
the physical trajectories of Fig.4.a ($H_0=1/24$) with that of the
trajectories of Fig.4.f. ($H_0=1/6$). These six displays are obtained by
plotting ``3PS's" over $(q_1,q_2,t)$ and changing the perspective,
appropriately, using the mouse:
\begin{verbatim}
> for ic in [ics[1/24][1..3],ics[1/6][1..3]] do
> poincare(H,t=-100..100,{ic},scene=[q1,q2,t],stepsize=.05,iterations=3,3):
> od:
\end{verbatim}

\dthreegraphic{2in}{2in}{f5a}{f5b}{f5c}
\noindent
\begin{tabular}{p{2in} p{2in} p{2in}}
{\footnotesize Fig.5.a. $H_0$=$1/24$, ics[1/24][1],
$H$-dev.=$2\x 10^{-7}\,\%$, time: $124 s.$}
&
{\footnotesize Fig.5.b. $H_0$=$1/24$, ics[1/24][2],
$H$-dev.=$2\x 10^{-7}\,\%$, time: $149 s.$}
&
{\footnotesize Fig.5.c. $H_0$=$1/24$, ics[1/24][3],
$H$-dev.=$3\x 10^{-7}\,\%$, time: $159 s.$}
\end{tabular}
\medskip

\dthreegraphic{2in}{2in}{f5d}{f5e}{f5f}
\noindent
\begin{tabular}{p{2in} p{2in} p{2in}}
{\footnotesize Fig.5.d. $H_0$=$1/6$, ics[1/6][1],
$H$-dev.=$4\x 10^{-7}\,\%$, time: $169 s.$}
&
{\footnotesize Fig.5.e. $H_0$=$1/6$, ics[1/6][2],
$H$-dev.=$5\x 10^{-7}\,\%$, time: $178 s.$}
&
{\footnotesize Fig.5.e. $H_0$=$1/6$, ics[1/6][2],
$H$-dev.=$6\x 10^{-7}\,\%$, time: $191 s.$}
\end{tabular}

\medskip
The contrast between Figs.5.(a,b,c) and Figs.5.(d,e,f) emphasizes the
connection between the smoothness of the curves over a surface-of-section
and the regularity of the corresponding physical trajectories.

\subsection{An example from general relativity}
Chaos in General Relativity is an issue of rich debate. Since the pioneering
work of Beliinski, Khalatnikov and Lifshitz\cite{BKL}, who discussed chaotic
behavior in anisotropic Bianchi IX models, much advance has been made.
Francisco and Matsas\cite{matsas}, studying that model, originally noted that
the Liapunov exponents tend to zero in the numerical experiments, due to the
choice made for the time variable. This result has been confirmed by several
authors\cite{several}. In this context, the Poincar\'e surfaces-of-section
play a crucial role, since the destruction of KAM surfaces does not depend on
the time variable and constitutes the signature of chaos.

Bearing the above remarks in mind, we examine here a numerical study appearing
in a recent work by Calzetta and El Hasi\cite{calzetta2}. The authors
developed a perturbative study of the influence of the scalar radiation field
on the expansion of the universe in the early stages of inflation. They
performed numerical experiments to exhibit chaotic behavior indicated by the
destruction of tori structures, formation of cantori, and Arnold diffusion.
The Hamiltonian for the model is given by:
\begin{equation}
H=\frac{1}{2}\,(-p_1^2-q_1^2+2\,\Lambda\,q_1^4+p_2^2+q_2^2+m^2\,q_1^2\,q_2^3)=0
\label{H3}
\end{equation}
Such an expression describes closed homogeneous and isotropic universes with a
cosmological constant, $\Lambda$, playing the role of the inflaton. The
degrees of freedom of the model are the ``radius of the universe", $q_1$, and
the conformally scaled radiation field, $q_2$; $m$ represents its mass, and
$p_1$, $p_2$ are the conjugated momenta.

Taking convenient values $m=0.65$ and $\Lambda=0.125$ and choosing the
$q_2=0$, $(q_1,p_1)$ plane as in \cite{calzetta2}, it is possible to reproduce
the relevant 2PS there displayed as Fig.2 as follows. First, an appropriate
set of one hundred lists of initial conditions, satisfying $H_0=0$ and in
accordance with the initial values for $p_1$, $q_1$ indicated in that paper,
is generated via:
\begin{verbatim}
> ic1 := gin(H,{t=0,p1=-.2..-0.7,q1=0,q2=0,energy=0},15):
> ic2 := gin(H,{t=0,p1=-.7..-0.812,q1=0,q2=0,energy=0},85):
> ics:= ic1 union ic2:
> poincare(H,t=0..300,ics,
>      stepsize=.1,iterations=3,scene=[q1=-1.5..1.5,p1=-1..1,q2=-1..1]);
\end{verbatim}
\dgraphic{5in}{3in}{f6}
\noindent
\begin{quote}
{\footnotesize Fig.6 The presence of smooth curves related to KAM surfaces, as
well as a region of broken tori. More than 10,000 intersection points,
absolute $H$-dev.$\approx$$10^{-8}$. Time: $35$ minutes.}
\end{quote}

Another good test is given by the plotting of a 3PS over $(p_1,q_1,q_2)$, for
initial conditions very close to the critical point $P_c$, where
$p_1=p_2=q_1=q_2=0$:

\begin{verbatim}
> ics := gin(H,{t=0,p1=0..-0.31,q1=0,q2=0,energy=0},40);
> poincare(H,t=0..20,ics,
>          stepsize=.1,scene=[q1=-.3..0.3,p1=-.3..0.3,q2=-.1..0.1],3);
\end{verbatim}

\dtwographic{3in}{2.8in}{f7a}{f7b}

\noindent
\begin{tabular}{p{3in} p{3in}}
{\footnotesize Fig.7.a. $\theta$=$-160$, $\phi$=$50$.
Regular circles close to $P_c$, in agreement with complex eigenvalues
for the linearized system. Time: $61 s.$}
&
{\footnotesize Fig.7.b. $\theta$=$-20$, $\phi$=$18$. The action of higher
order terms in splitting-up the KAM surfaces. Absolute $H$-deviation
$<10^{-8}$.
}
\end{tabular}

\section{Conclusions}

This work presented a set of software-tools for the plotting of Poincar\'e
surfaces-of-section of Hamiltonian systems. This method is a valuable tool for
studying dynamical systems, since it conveys relevant information on the
dynamics, in a simple manner. Due to the characteristic flexibility of symbolic
programming languages, the main result was an easy-to-use package of commands
permitting reasonably fast and varied numerical studies of Hamiltonian
systems, in a general purpose symbolic computing environment.

An important remark, taking into account that this software was written for
realizing intensive numerical studies, is that great emphasis was put in the
{\it interactive} character of the package. That is, the user is given the
possibility of alternatively searching for ``first order" phenomena at
remarkably high speed, obtaining draft Poincar\'e sections in just a few
seconds; or ``high order" detailed 2D/3D projections, displaying ``islands"
and the inner structure of a PS, as desired.

On the other hand, this is a first version of the package and, as such, it
does not make use of the theory of Liapunov coefficients, does not discuss the
determination of the analytical Poincar\'e mappings, does not feature options
related to the structure of Arnold's web, and is not designed to study the
interesting case of dissipative Hamiltonians and correlated phenomena (strange
attractors etc.). All these topics are possible extensions of this work and we
expect to report related work in the near future.


\begin{thebibliography}{99}

\bibitem{lichtenberg} A.J. Lichtenberg and M.A. Lieberman, {\it Regular and
Stochastic Motion}, Applied Mathematical Sciences {\bf 38} (Springer Verlag,
New York, 1994).

\bibitem{henonheiles} M. H\'enon and C. Heiles, The Astronomical Journal, {\bf
69} (1963) 73.

\bibitem{calzetta2} E. Calzetta and C. El Hasi, Phys. Rev. D {\bf 51} (1995)
2713.

\bibitem{PDEtools} E.S. Cheb-Terrab and K. von B\"ulow, Comp. Phys. Comm., 90
(1995) 102-116, and ``Analytical Solving of Partial Differential
Equations using Symbolic Computing", to appear in the {\it Proceedings of
Computing in High Energy Physics} (Rio de Janeiro, CHEP95, World Scientific
Publishing Co., 1995).

\bibitem{ford} J. Ford, S.D. Stoddard and J.S. Turner, Prog. Theor. Phys {\bf
50} (1973) 1547.

\bibitem{henon} M. H\'enon, Phys. Rev. B {\bf 9} (1974) 1925.

\bibitem{BKL} V.A. Belinskii, E.M. Lifshitz and I.M. Khalatnikov, Sov. Phys.
Usp., {\bf 13} (1971) 745.

\bibitem{matsas} G. Francisco and G.E.A. Matsas, Gen. Relat. Grav., {\bf 20}
(1988) 1047.

\bibitem{several} A. Burd and R. Tavakol, Phys. Rev. D {\bf 47} (1993) 5336,
and references therein.

\end{thebibliography}
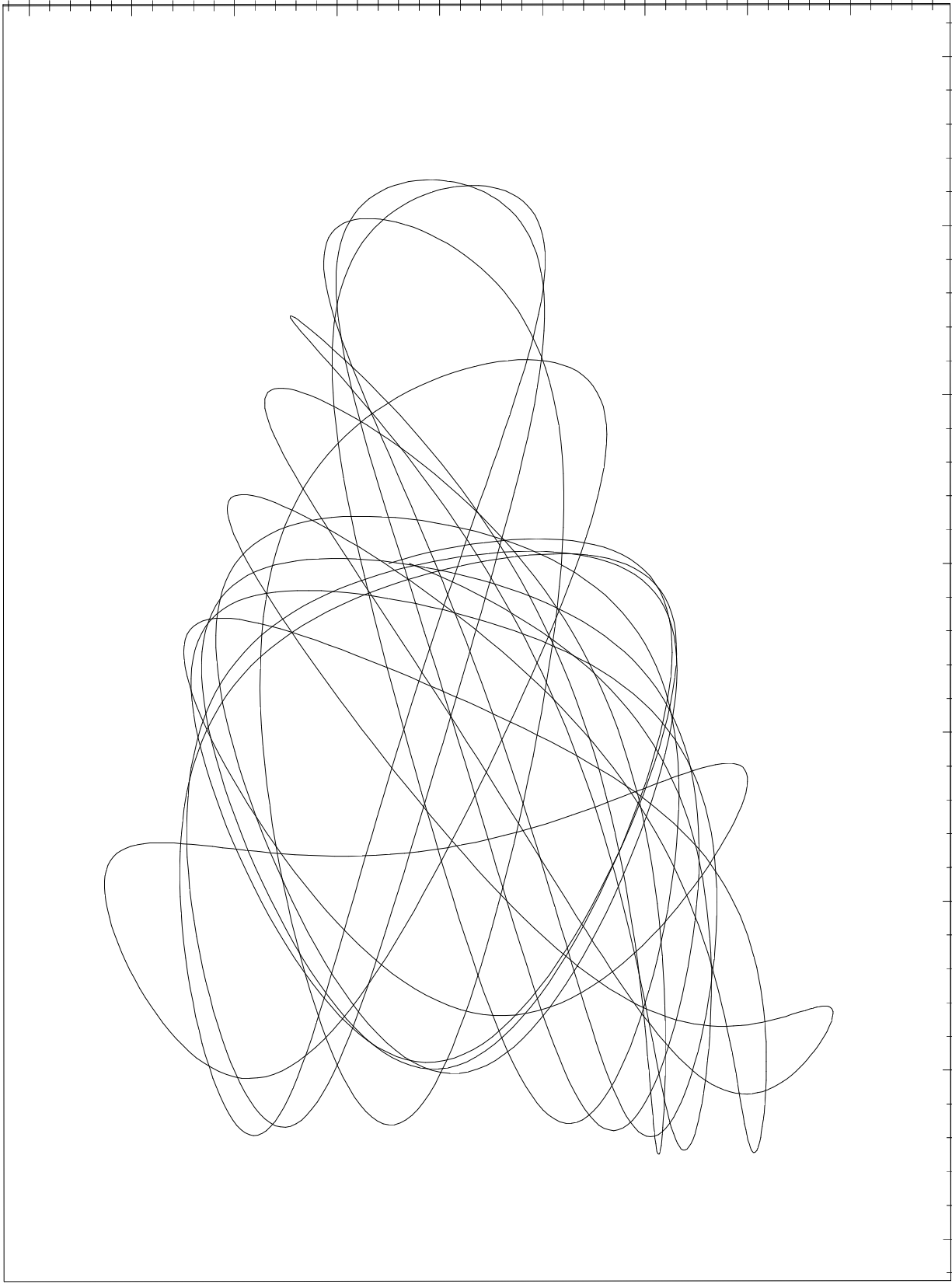
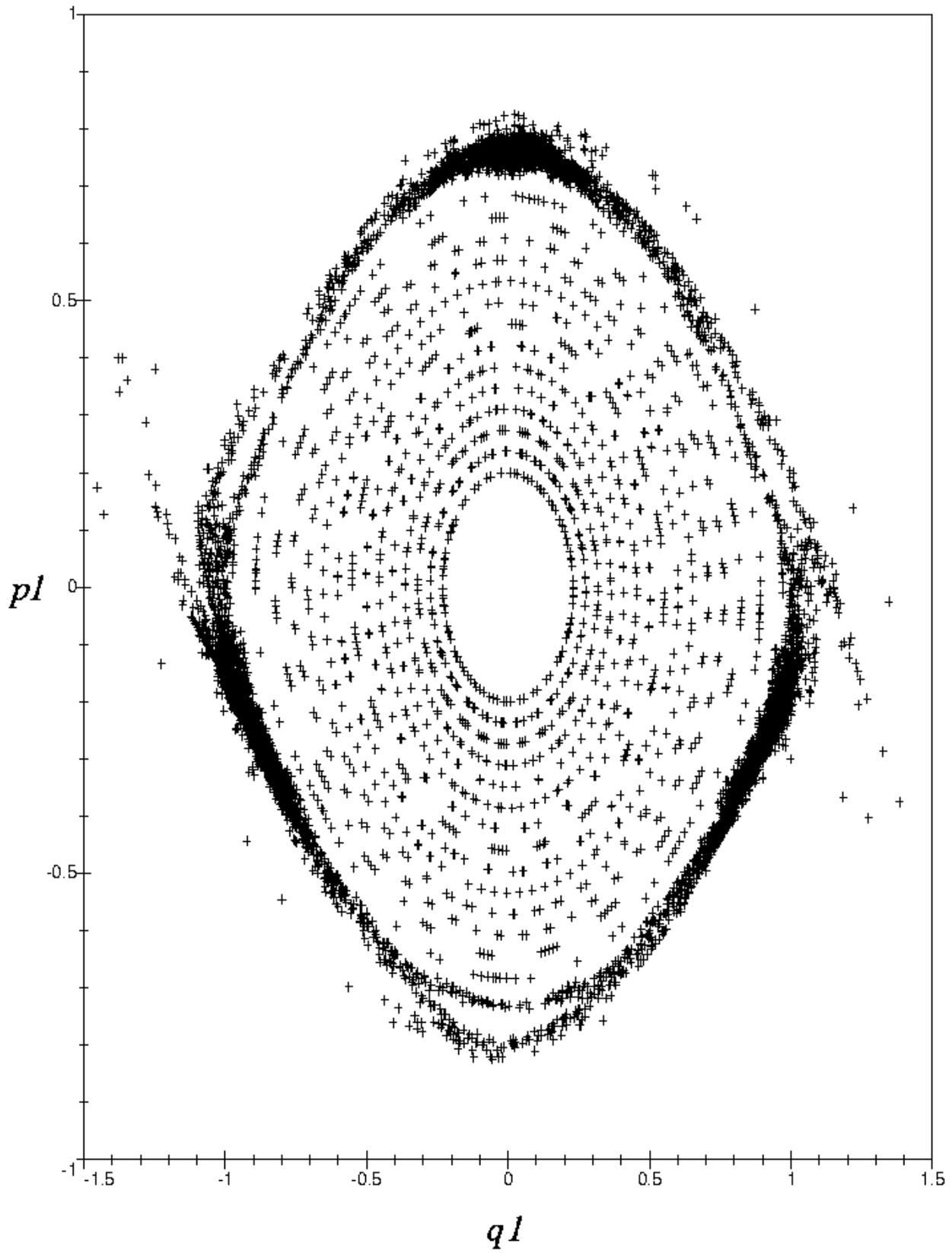
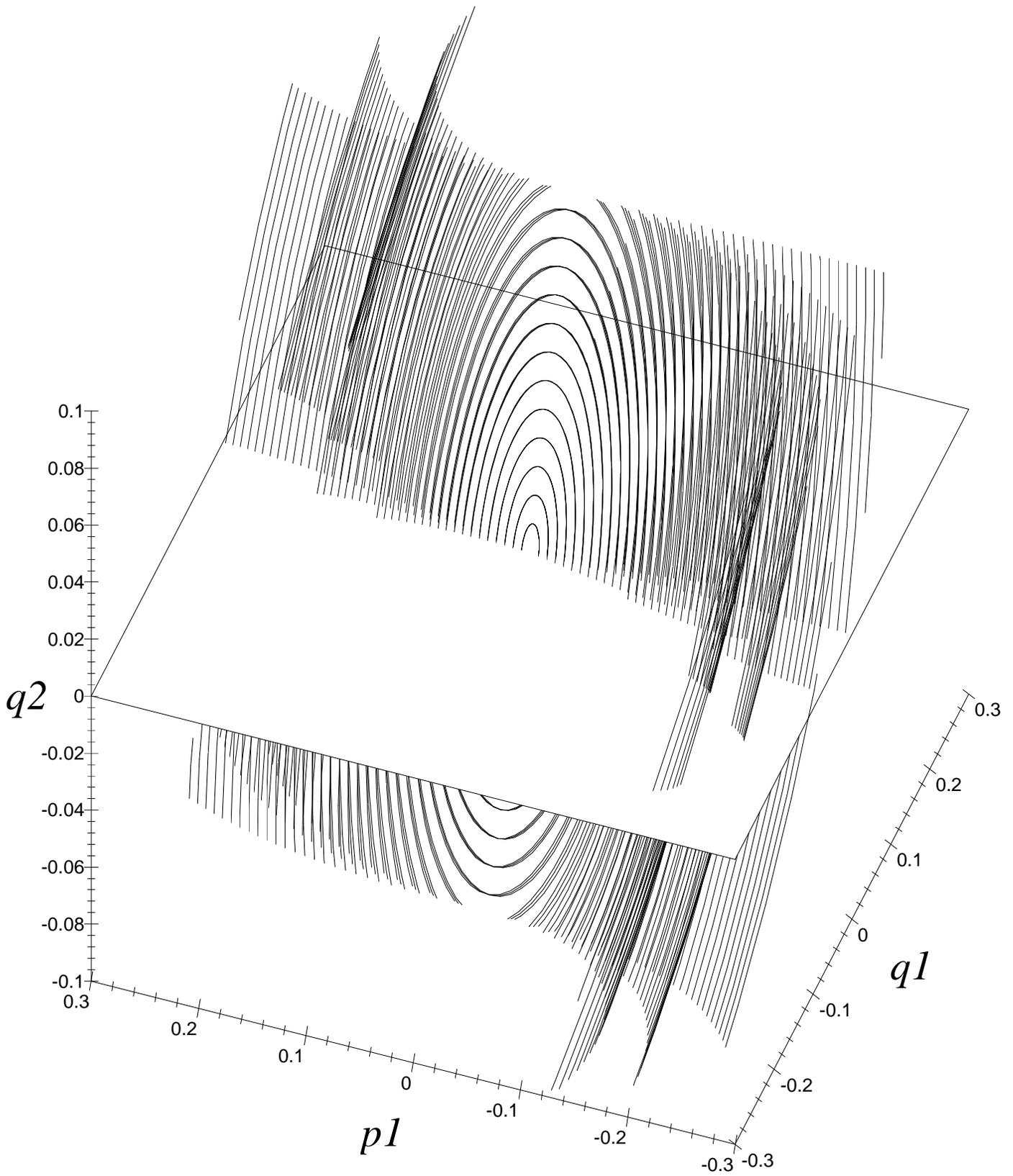
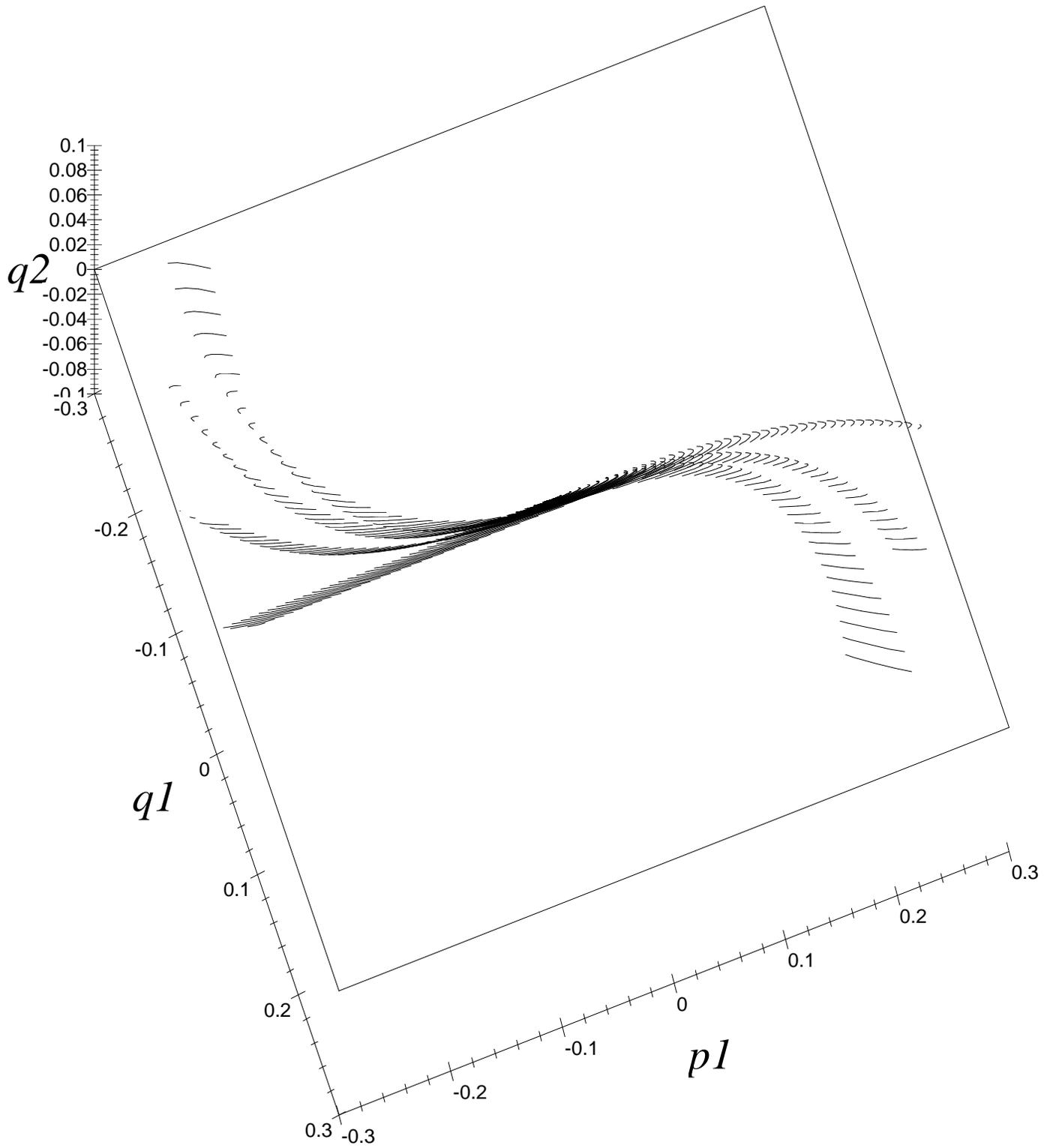
\end{document}